%% file: main_final.tex
\documentclass[conference,10pt]{IEEEtran}

\usepackage{amsmath, mathrsfs, mathtools}
\usepackage{amsfonts}
\usepackage{amssymb}
\usepackage{color}
\usepackage{multirow}
\usepackage{stmaryrd}
\usepackage{yfonts}
\usepackage{mathabx}
\usepackage{caption}
\usepackage{subcaption}
\usepackage{float}
\usepackage{stfloats}
\usepackage{amsfonts}
\usepackage{tcolorbox}
\usepackage[numbers,sort&compress]{natbib}
\usepackage{multicol}
\usepackage{algorithm}
\usepackage{algpseudocode}
\usepackage{pifont}
\usepackage{varwidth}
\usepackage{afterpage}
\usepackage{balance}
\usepackage{lipsum}
\usepackage{tikz, pgfplots,graphicx,xcolor}
\usetikzlibrary{plotmarks,spy,backgrounds}
\pgfplotsset{compat=newest}

\input{symbols.tex}

\input{macros.tex}

\usepackage{textcomp}

\usepackage{geometry}
\allowdisplaybreaks
\title{Near-Field Imaging by Exploiting Frequency Correlation in  Wireless Communication Networks}
\author{Tianyu Yang, Kangda Zhi, Shuangyang Li, and Giuseppe Caire\\
Faculty of Electrical Engineering and Computer Science,  Technical University of Berlin, Berlin, Germany
\\
	$\{$tianyu.yang, kangda.zhi, shuangyang.li, caire$\}$@tu-berlin.de

\vspace{-20pt}
}

\begin{document}
\maketitle
\begin{abstract}
    In this work, we address the near-field imaging under a wideband wireless communication network by exploiting both the near-field channel of a uniform linear array (ULA) and the image correlation in the frequency domain. We first formulate the image recovery as a special multiple measurement vector (MMV) compressed sensing (CS) problem, where at various frequencies the sensing matrices can be different, and the image coefficients are correlated. To solve such an MMV problem with various sensing matrices and correlated coefficients, we propose a sparse Bayesian learning (SBL)-based solution to simultaneously estimate all image coefficients and their correlation on multiple frequencies. Moreover, to enhance estimation performance, we design two illumination patterns following two different criteria. From the CS perspective, the first design minimizes the total coherence of the sensing matrix to increase the mutual orthogonality of the basis vectors. Alternatively, to improve SNR, the second design maximizes the illumination power of the imaging area. Numerical results demonstrate the effectiveness of the proposed SBL-based method and the superiority of the illumination designs.   
\end{abstract}


\section{Introduction}
By leveraging the increasing frequency bands and antenna array size and sharing the same hardware and time-frequency resource between communication and sensing functionalities, integrated sensing and communication (ISAC) has been listed as a major usage scenario in 6G \cite{wp5d2023m}. 
As a promising branch of ISAC, integrated imaging and communication (IIAC) focuses on 
wireless imaging \cite{li2021lightweight, yang2025illumination}, which has a huge potential to provide secure environmental imaging acquisition, enabling crucial applications such as security inspections, digital twin, and environmental reconstruction that can be used to enhance communication performance, e.g., providing side information for beam selection and beam-user association.

Different from the ``target sensing'' that focuses on target presence/absence detection and radar-style target parameter estimation (e.g., range, velocity, angle of arrival (AoA)), IIAC considers the recovery of scattering coefficients in a continuous imaging area.   
Although the task of imaging recovery based on wireless signal has been widely studied in the radio-frequency (RF) imaging community, e.g., 
microwave imaging \cite{shao2020advances}, diffraction tomography \cite{ren20183}, and
synthetic aperture radar (SAR) \cite{fang2013fast}, their methods 
rely on all observations from each transmit antenna element to each receive antenna element, requiring 
signal-space orthogonality of the probing signals sent by each transmit antenna. This signaling approach is not suitable for IIAC, where the transmitter(s) usually apply beamforming (BF) to serve communicating users in space division multiplexing, and the imaging sensors must cope with these signaling formats.

In recent IIAC works, a least-squares (LS)-based method is used to recover the image from the noisy observations 
\cite{torcolacci2024holographic, yang2025illumination}. However, the assumption under such pseudoinverse approaches is that the image dimension is not larger than the observation dimension, which strongly restricts the image resolution. Additionally, it is well known that the pseudoinverse approach suffers severely from the off-grid error, i.e., the error due to that the true pixels are not exactly located on the discrete grid used by the algorithm.

In this work, we focus on the imaging-centric IIAC problem under a near-field wideband channel by exploiting both the image spatial sparsity and frequency correlation. Leveraging the dual spatial resolution (angle-distance) in the near-field channel, we recover a 2D image using only a uniform linear array (ULA). 
In contrast to the LS approaches, we formulate the image recovery problem as a special multiple measurement vector (MMV) compressed sensing (CS) problem, where the sensing matrices for the unknown coefficients at different frequencies can be different. In order to handle a more general problem with various but correlated unknown coefficients in a wideband system, we use a sparse Bayesian learning (SBL)-based method to jointly estimate all coefficients and their correlation matrix. Additionally, to enhance the imaging performance, we also optimize the illumination BF pattern to minimize the total coherence of the sensing matrix and maximize the worst signal-to-noise ratio (SNR).

\section{System model}

\begin{figure}
    \centering
    \includegraphics[width=0.8\linewidth]{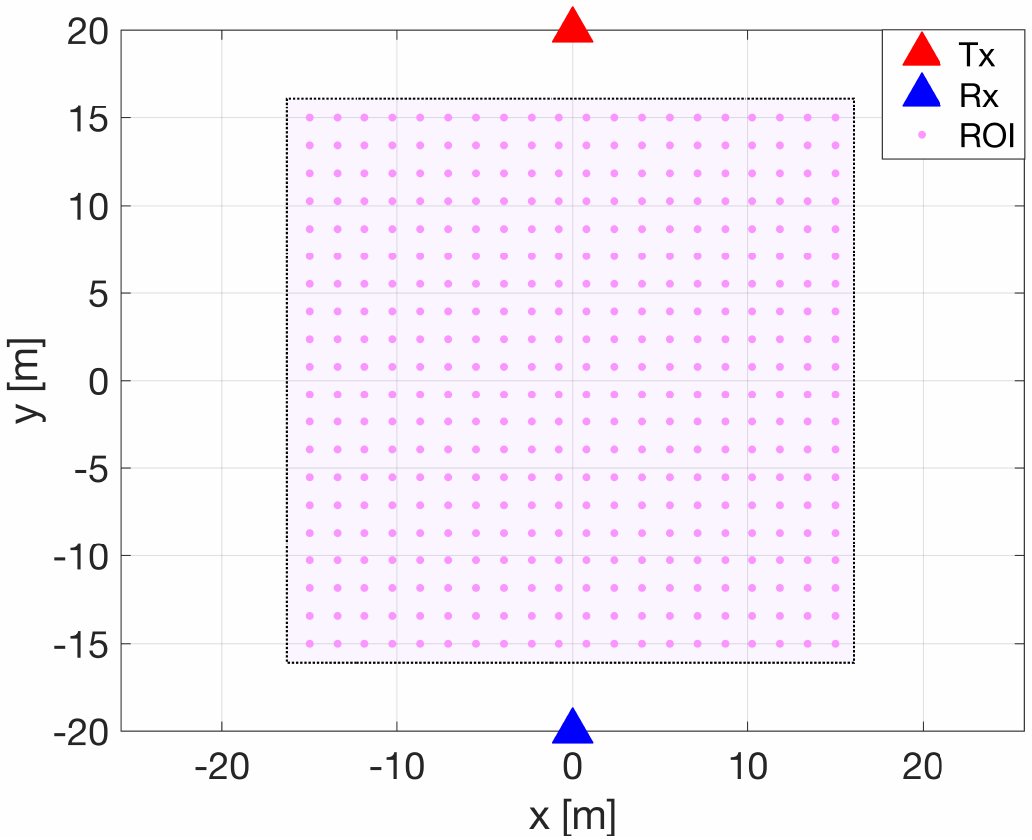}
    \caption{Considered imaging-centric near-field IIAC with one transmitter (Tx) and one receiver (Rx) equipped with ULA to recover image in the near-field region of interest (ROI). }
    \label{fig:IIAC}
    \vspace{-5mm}
\end{figure}

We consider a bistatic IIAC scenario\footnote{Note that our method is applicable for monostatic, bistatic, and even multistatic cases. We emphasize that the monostatic sensing with co-located transmitter and receiver requires the capability of full-duplex communication, which is still restricted in realistic communication systems due to unavoidable residual self-interference.}, where the transmitter (Tx) and receiver (Rx) are respectively located at $\rv_{\rm t}$ and $\rv_{\rm r}$ according to a global 2D coordinate, see Fig.~\ref{fig:IIAC}.    
The wideband communication system makes use of OFDM signaling at a central frequency $f_{ c}$ with subcarrier spacing $\Delta f$, and number of subcarriers $N$.
Given the region of interest (ROI) for imaging with its coordinate set $\mathcal{A}$, the effective imaging channel at subcarrier $n$ is the integral over the ROI as 
\begin{align}\label{eq:channel_model}
    \Hm_n  &= \int_{ \mathcal{A}} \rho_{n}(\rv) \eta(\rv) \av(\rv) \bv^\herm(\rv) t_n(\rv) d\rv, \;  n\in[N],
\end{align}
where $\rv \in \mathcal{A}$ is the coordinate of a position in the ROI $\mathcal{A}$, $\rho_n(\rv)$ is the complex reflection coefficient at position $\rv$, $\eta(\rv)$ is the effective pathloss from the Tx to the Rx through the position $\rv$, $\av(\rv)$ and $\bv(\rv)$ are angular steering vectors, and $t_n(\rv)$ is the phase shift due to the delay at subcarrier $n$. Specifically, assuming that the Tx and Rx are respectively equipped with $M_{\rm t}$ and $M_{\rm r}$ antennas in a fully digital uniform linear array (ULA), the $m$-th elements of the near-field angular steering vectors $\av(\rv)$ and $\bv(\rv)$ are given as
\begin{align}
    [\av(\rv)]_m &= e^{-j \frac{2\pi}{\lambda_{ c}} d_{\rm r}^m(\rv)  },\quad \forall m \in [M_{\rm r}], \\ 
    [\bv(\rv)]_m &= e^{-j \frac{2\pi}{\lambda_{ c}} d_{\rm t}^m(\rv)  },\quad \forall m \in [M_{\rm t}],
\end{align}
where $\lambda_{c} = \frac{c_0}{f_{ c}}$ is the carrier wavelength with $c_0$ being the speed of light, and where 
$d_{\rm t}^m(\rv) = \|\rv_{\rm t}^m - \rv \|_2$ and  $d_{\rm r}^m(\rv) = \|\rv_{\rm r}^m - \rv \|_2$  are the distance between the position $\rv$ and $m$-th antenna element of Tx and Rx with $\rv^m_{\rm t}$ and $\rv^m_{\rm r}$ being the positions of $m$-th antenna element of Tx and Rx, respectively.  Moreover, 
the phase shift $t_n(\rv)$ is given as\footnote{We assume that the phase shift due to delay is the same for all antenna pairs. This assumption holds whenever the array aperture is not comparable to the Tx-Rx distance, see e.g., \cite{cui2022channel}. Please also refer to the next footnote.} 
\begin{align}
    t_n(\rv) &= e^{-j2\pi (n-1) \tau(\rv) \Delta f},
\end{align}
where $\tau(\rv) = \frac{d_{\rm t}(\rv) + d_{\rm r}(\rv)}{c_0}  $ is the delay from Tx to Rx reflected by the position $\rv$ with $d_{\rm t} = \|\rv_{\rm t}-\rv\|_2$ $(d_{\rm r} = \|\rv_{\rm r}-\rv\|_2)$ being the distance between the position $\rv$ and Tx (Rx).
Similarly, the effective pathloss
$\eta(\rv)$ is given as  \cite{richards2005fundamentals}\footnote{Instead of antenna element-dependent pathloss, we use approximate array position-dependent pathloss that is the same for all antenna elements from a scatter point. Note that under half-wavelength antenna spacing and 50 GHz carrier frequency, the antenna spacing is only 3 mm, resulting in a 0.3 m antenna aperture of 100 antennas, which is negligible compared to the Tx-Rx distance. Additionally, our method is not limited by the approximate pathloss model and can be directly applied to the problem with antenna element-dependent pathloss. } 
     $\eta(\rv) = \sqrt{\frac{\lambda^2}{(4\pi)^3 d^2_{\rm t} d^2_{\rm r}}}$.
Finally, we assume that the complex reflection coefficients $\rho_n(\rv)$ of different positions are independent, and of the same position but different subcarriers are correlated, which is mathematically given as 
\begin{align}\label{eq:correlation}
    \bE[\rho_n(\rv) \rho_{n'}^*(\rv')] = \gamma(\rv) [\Psim]_{n,n'} \delta(\rv - \rv'), \; \rv\in \mathcal{A}, n\in[N],
\end{align}
where $\delta(\cdot)$ is the Dirac delta function and $\Psim \in \bC^{N\times N}$ is a positive definite matrix capturing the correlation of $N$ subcarriers, in which we assume that the coefficient correlations in frequency of all positions are all the same as $\Psim$. Moreover, $\gamma(\rv)$ is the radar cross section (RCS) of the position $\rv$.

We consider an IIAC problem, where the Tx uses the communication data signals to simultaneously serve communication users and process RF imaging. We assume that imaging Rx knows the data signal and thus can use it for imaging. We focus on an imaging-centric case, where we do not optimize the communication performance. Assuming a normalized data signal and given the transmit BF vector at subcarrier $n$ as $\xv_n$, the received signal at subcarrier $n$ is given as
\begin{align}
    \yv_n = \Hm_n \xv_n + \nv_n,
\end{align}
where $\sum^N_{n=1}\|\xv_n\|^2_2 \leq P_{\rm tot}$ with $P_{\rm tot}$ being the total transmit power, and $\nv_n\sim \mathcal{CN}(\mathbf{0}, N_0\Id_{M_{\rm r}})$ is the additive white Gaussian noise (AWGN) with noise power $N_0$.

Given the received signal of all $N$ subcarriers as $\Ym = [\yv_1, \dots, \yv_N]$, the imaging task is to estimate the reflection coefficients $\{\rho_n(\rv)| n\in[N], \rv\in\mathcal{A}\}$. Note that normally only the strength of the imaging coefficients is considered for evaluation of the imaging quality.

\section{SBL-based Wideband Imaging}
Our scheme follows the idea of parametric representation of the imaging channel and estimates the parameters using SBL with the exploitation of their frequency correlation. First, we discretize the ROI $\mathcal{A}$ into $Q$ square cells whose central positions are given as $\{\rv_i \in \mathcal{A}| i \in [Q]\}$. Similar to \cite{yang2025illumination}, we assume that each cell performs a perfect isotropic reflection characterized by a reflection coefficient $\widetilde{\rho}_n(\rv_i)$. Thus, the imaging channel can be written in a linear form as
\begin{align}
    \Hm_n &\approx \sum^Q_{i = 1} \widetilde{\rho}_n(\rv_i) \eta(\rv_i) t_n(\rv_i) \av(\rv_i)\bv(\rv_i)^\herm,\\
    &=\Am \diag( \etav)\diag(\underbrace{\widetilde{\rhov}_n \odot \tv_n }_{\triangleq \uv_n} )  \Bm,\\
    & = \Am \diag( \etav)\diag(\uv_n) \Bm,
\end{align}
where $\widetilde{\rhov}_n = [\widetilde{\rho}_n(\rv_1), \dots, \widetilde{\rho}_n(\rv_Q)]^\transp \in \bC^Q $ and $\etav = [\eta(\rv_1),\dots,\eta(\rv_Q)]^\transp \in \bC^Q$ are respectively the RCS-related reflection coefficients and effective pathloss of $Q$ cells. $\Am = [\av(\rv_1), \dots, \av(\rv_Q)] \in \bC^{M_{\rm r}\times Q}$ and $\Bm = [\bv(\rv_1), \dots, \bv(\rv_Q)]^\herm \in \bC^{Q \times M_{\rm t}}$ are the steering matrices. $\tv_n = [t_n(\rv_1), \dots, t_n(\rv_Q)]^\transp  \in \bC^Q$ is the delay response vector at subcarrier $n$. Combining the coefficients and delay response vectors, we obtain the vector $\uv_n$ at the subcarrier $n$.

Then, the noisy received signal can be approximated as 
\begin{align}
    \yv_n &\approx \Am \diag( \etav)\diag(\uv_n) \Bm \xv_n + \nv_n,\\
    &=\underbrace{\Am \diag( \etav) \diag(\Bm \xv_n)}_{\triangleq\Phim_n } \uv_n + \nv_n,\\
    &=\Phim_n \uv_n + \nv_n.
\end{align}
The noisy observations of all subcarriers can be written as 
\begin{align}\label{eq:observation}
    \Ym \approx  [\Phim_1 \uv_1, \dots, \Phim_N \uv_N] + \Nm,
\end{align}
where $\Nm = [\nv_1, \dots, \nv_N]$. We further define the coefficient matrix $\Um \triangleq [\uv_1, \dots, \uv_N]$ and note that $\Um$ can be written as
\begin{align}
    \Um = \Pm  \odot \Tm \in \bC^{Q \times N},
\end{align}
where $\Pm = [\widetilde{\rhov}_1, \dots, \widetilde{\rhov}_N]$ and  $   \Tm = [\tv_1, \dots, \tv_N]$. 

Assuming that the image in ROI is sparse, i.e., $\widetilde{\rhov}_n$ is a sparse vector, we have that all columns of $\Um$ are sparse vectors with a common sparse support. Estimating the unknown $\Um$ from the observation $\Ym$ is an MMV problem in CS, where the sensing matrices $\Phim_n$ for different observations might be different, relating to the design of $\{\xv_n\}$.
We use the SBL approach to solve such an MMV problem with varying sensing matrices and exploitation of coefficient correlation.
The SBL treats the unknown quantities as random variables that follow some prior distribution. From \eqref{eq:correlation}, we can assume that the reflection coefficients of different cells are independent, and of the same cell but different subcarriers are correlated, i.e.,
\begin{align}
    \bE[\widetilde{\rho}_n(\rv_i) \widetilde{\rho}_{n'}^*(\rv_{i'})] = \gamma_i [\Psim]_{n,n'} \mathbb{1}_{i = i'}, \; i\in[Q], n\in[N],
\end{align}
where $\mathbb{1}_{i=i'}$ is the indicator function of the condition $i=i'$.
Further defining the parameters $\Gammam = \diag(\gammav) = \diag([\gamma_1, \dots, \gamma_Q]^\transp) \in \bR^{Q\times Q}$, we assume that the prior distribution of $\Um$ is given as
\begin{align}
    p(\Um; \Gammam, \Psim) = \prod^Q_{i=1}\Big( p(\bar{\uv}_i; \gamma_i, \Psim) \sim \mathcal{CN} (\mathbf{0},\Zm_i)\Big),
\end{align}
where $\bar{\uv}_i \triangleq ([\Um]_{i,:})^\transp = \diag(\bar{\tv}_i)\bar{\rhov}_i$ with $\bar{\tv}_i \triangleq ([\Tm]_{i, :})^\transp$ and $\bar{\rhov}_i \triangleq ([\Pm]_{i, :})^\transp, \; i \in [Q]$. Moreover, we have 
\begin{align}
    \Zm_i &\triangleq \bE\left[\bar{\uv}_i \bar{\uv}_i^\herm\right] =  \diag(\bar{\tv}_i)\bE\left[\bar{\rhov}_i \bar{\rhov}_i^\herm\right] \diag(\bar{\tv}_i)^\herm \\
    &=\gamma_i \underbrace{ \diag(\bar{\tv}_i)\Psim \diag(\bar{\tv}_i)^\herm}_{\triangleq \Wm_i} = \gamma_i \Wm_i.
\end{align}
Then, the SBL approach maximizes the Bayesian evidence $p(\Ym;\Gammam, \Psim)$ to infer the parameters $\gammav$ and $\Psim$, and then estimates $\Um$ via the posterior mean of the posterior distributions $p(\Um | \Ym; \gammav, \Psim)$. 
Hence, the cost function of SBL is the minus log-likelihood function for maximum-likelihood (ML) estimation of the parameter vector $\gammav$ and $\Psim$, which is given as
\begin{align}\label{eq:cost_function}
    -\log p(\bar{\yv}; \Gammam, \Psim) = \log(\det(\Sigmam_{\bar{\yv}})) + \bar{\yv}^\herm \Sigmam_{\bar{\yv}}^{-1}\bar{\yv},
\end{align}
where $\bar{\yv} \triangleq \vec(\Ym)$, $\Sigmam_{\bar{\yv}}$ is the covariance matrix of $\bar{\yv}$, and the constant term is ignored. The cost function in \eqref{eq:cost_function} is not a convex function, which is normally solved iteratively using the expectation–maximization (EM) method. 

Denoting the estimated parameters in the $\ell$-th iteration as $\widehat{\Gammam}^{(\ell)} = \diag(\widehat{\gammav}^{(\ell)}) = \diag([\widehat{\gamma}_1^{(\ell)}, \dots, \widehat{\gamma}_Q^{(\ell)}]^\transp) $ and $\widehat{\Psim}^{(\ell)}$, and treating $\Um$ as hidden variables, each E-step of the EM method evaluates the average minus log-likelihood of the complete data set $\{\Ym, \Um\}$ as\footnote{In this work, we assume that the noise power $N_0$ is known. However, we emphasize that the proposed SBL approach can also be applied when the noise power is unknown, in which the noise power will be jointly estimated with other unknown parameters. } 
\begin{align}
&\mathcal{L}^{(\ell)}\left(\gammav, \Psim|\widehat{\gammav}^{(\ell)}, \widehat{\Psim}^{(\ell)}\right)\notag  \\ 
&\!\!\!= 
\bE_{\Um|\Ym; \widehat{\gammav}^{(\ell)}, \widehat{\Psim}^{(\ell)}}[-\log p(\Ym, \Um;\gammav, \Psim) ]\\
    &\!\!\!=\bE_{\Um|\Ym; \widehat{\gammav}^{(\ell)}, \widehat{\Psim}^{(\ell)}}\left[\underbrace{-\log p(\Ym|\Um)}_{\text{constant due to known $N_0$}} \underbrace{-\log p(\Um;\Gammam, \Psim)}_{\text{using prior of $\Um$}}\right] \\
    &\!\!\!\overset{(a)}{=} \bE_{\Um|\Ym; \widehat{\gammav}^{(\ell)}, \widehat{\Psim}^{(\ell)}} \left[\sum^Q_{i=1} -\log \left(\frac{\exp\left(-\bar{\uv}_i^\herm \Zm_i^{-1}\bar{\uv}_i\right)}{\pi^N \det( \Zm_i)} \right)\right]\\
    &\!\!\!\overset{(b)}{=}\sum^Q_{i=1} \log(\det(\Zm_i)) + \bE_{\bar{\uv}_i | \Ym; \widehat{\gamma}_i^{(\ell)}, \widehat{\Psim}^{(\ell)} }\left[\bar{\uv}_i^\herm \Zm_i^{-1}\bar{\uv}_i\right]\\
    &\!\!\!=\sum^Q_{i=1} N \log(\gamma_i) + \log\det(\Wm_i) + \bE_{\bar{\uv}_i | \Ym; \widehat{\gamma}_i^{(\ell)}, \widehat{\Psim}^{(\ell)}}\left[\bar{\uv}_i^\herm \Zm_i^{-1}\bar{\uv}_i\right] \label{eq:L_ini}
\end{align}
where in $(a)$ and $(b)$ the constant terms that do not contribute to the subsequent optimization of the parameters are dropped.

To calculate the last term in \eqref{eq:L_ini}, we need a posterior density of each $\bar{\uv}_i$, which can be obtained from the posterior density of the full $\Um$. Concretely, we define the vectorization of the transpose of $\Ym$ as
\begin{align}
    \widetilde{\yv} &\triangleq     \vec\left(\Ym^{\transp}\right)\\
    &=\underbrace{\left(\sum_{n=1}^N \Phim_n \otimes \boldsymbol{\Upsilon}_n\right)}_{\triangleq \widetilde{\Phim}}\underbrace{\vec\left(\Um^\transp\right)}_{\triangleq \widetilde{\uv}} + \underbrace{\vec\left(\Nm^\transp\right)}_{\triangleq \widetilde{\nv}} \\
    &= \widetilde{\Phim} \widetilde{\uv} +\widetilde{\nv},\label{eq:linear_form}
\end{align}
where $\boldsymbol{\Upsilon}_n \in \{0,1\}^{N\times N}$ is a $N$ by $N$ square matrix whose elements are all zeros except $[\boldsymbol{\Upsilon}_n]_{n,n} = 1$.\footnote{When the BF vectors in all subcarriers are the same, we have the same $\Phim$ for all subcarriers and the corresponding $\widetilde{\Phim}$ is simplified as $\widetilde{\Phim} = \Phim \otimes \Id_N$.}  It is noticed that $\widetilde{\uv} = [\bar{\uv}_1^\transp, \dots, \bar{\uv}_Q^\transp]^\transp \in\bC^{NQ}$ is the stacking vector of all rows of $\Um$. The prior distribution of $\widetilde{\uv}$ can be rewritten as $p(\widetilde{\uv};\Gammam,\Psim) \sim \mathcal{CN} (\mathbf{0}, \widetilde{\Gammam})$, where $ \widetilde{\Gammam} = \text{blkdiag}(\Zm_1, \dots, \Zm_Q)$.
Then, given \eqref{eq:linear_form}, the posterior distribution of  $\widetilde{\uv}$ is $p(\widetilde{\uv}|\widetilde{\yv};\widetilde{\Gammam}) \sim \mathcal{CN}(\muv_{\widetilde{\uv}}, \Sigmam_{\widetilde{\uv}})$, where the posterior mean and posterior covariance at the $\ell$-th iteration using estimated parameters $\widehat{\gammav}^{(\ell)}$ and $\widehat{\Psim}^{(\ell)}$ are given as \cite{zhang2013extension}
{\small
\begin{align}
    \Sigmam_{\widetilde{\uv}}^{(\ell)} &= \left(\frac{1}{N_0}\widetilde{\Phim}^\herm \widetilde{\Phim} + \left(\widetilde{\Gammam}^{(\ell)}\right)^{-1}\right)^{-1} \in \bC^{NQ\times NQ}\\
    &=\widetilde{\Gammam}^{(\ell)} - \underbrace{\widetilde{\Gammam}^{(\ell)} \widetilde{\Phim}^\herm \left(N_0 \Id_{NM_{\rm r}} + \widetilde{\Phim} \widetilde{\Gammam}^{(\ell)} \widetilde{\Phim}^\herm \right)^{-1}}_{\triangleq \widetilde{\Am}^{(\ell)}} \widetilde{\Phim} \widetilde{\Gammam}^{(\ell)} \label{eq:poster_cov}\\
    \muv_{\widetilde{\uv}}^{(\ell)} &= \frac{1}{N_0} \Sigmam_{\widetilde{\uv}}^{(\ell)}\widetilde{\Phim}^\herm \widetilde{\yv} =\widetilde{\Am}^{(\ell)} \widetilde{\yv}. \label{eq:poster_mean}
\end{align}
}
Having \eqref{eq:poster_cov} and \eqref{eq:poster_mean}, we can calculate the last term in \eqref{eq:L_ini}, which is given as
{\small
\begin{align}
    &\bE_{\bar{\uv}_i|\Ym; \widehat{\gamma}^{(\ell)}_i, \widehat{\Psim}^{(\ell)}}\left[\bar{\uv}_i^\herm \Zm_i^{-1}\bar{\uv}_i\right] \\     &=\bE_{\bar{\uv}_i|\Ym; \widehat{\gamma}^{(\ell)}_i, \widehat{\Psim}^{(\ell)}}\left[\trace\left(\bar{\uv}_i\bar{\uv}_i^\herm \Zm_i^{-1}\right)\right]\\
    &=\trace\left(\bE_{\bar{\uv}_i|\Ym; \widehat{\gamma}^{(\ell)}_i, \widehat{\Psim}^{(\ell)}}[\bar{\uv}_i \bar{\uv}_i^\herm] \Zm_i^{-1}\right) \\
    &=\trace\left((\underbrace{\bar{\muv}_i^{(\ell)} (\bar{\muv}_i^{(\ell)})^\herm + \bar{\Sigmam}_i^{(\ell)} }_{\triangleq \Rm_i^{(\ell)}})\Zm_i^{-1}\right) \label{eq:E_u_bar},
\end{align}
}
where using MATLAB index notation we have
\begin{align}
    \bar{\muv}_i^{(\ell)} &\triangleq [\muv^{(\ell)}_{\widetilde{\uv}}]_{(i-1)N + 1: i N}, \quad i \in [Q] \\
    \bar{\Sigmam}^{(\ell)}_i &\triangleq [\Sigmam_{\widetilde{\uv}}^{(\ell)}]_{(i-1)N + 1: i N, (i-1)N + 1: i N}, \quad i \in [Q]
\end{align}
Putting \eqref{eq:E_u_bar} back to \eqref{eq:L_ini}, we have the cost function
\begin{align}
    \mathcal{L}^{(\ell)} &= \sum^Q_{i=1} N \log(\gamma_i) + \log \det\left(\Wm_i\right) + \trace\left(\Rm_i^{(\ell)}\Zm_i^{-1}\right).
\end{align}

In each M-step, the parameters $\gammav$ and $\Psim$ are updated by minimizing the cost function $\mathcal{L}^{(\ell)}$. Specifically, by setting the partial derivative of the cost function with respect to each $\gamma_i$ to zero, we obtain the update rule for $\gammav$, which is given as\footnote{$a\overset{!}{=}b$ means that $a$ shall be equal to $b$.}
\begin{align}
    \frac{\partial \mathcal{L}^{(\ell)}}{\partial \gamma_i} & = \frac{N}{\gamma_i} - \frac{\trace\left(\Rm_i^{(\ell)}\Wm_i^{-1}\right)}{\gamma_i^2} \overset{!}{=}  0 \\ 
    \Rightarrow \quad \widehat{\gamma}_i^{(\ell + 1)} & = \frac{1}{N}\trace\left(\Rm_i^{(\ell)}(\Wm_i^{(\ell)})^{-1} \right), \quad i\in[Q], 
\end{align}
where $\Wm^{(\ell)}_i$ uses the estimation $\widehat{\Psim}^{(\ell)}$.  
Similarly, by setting the gradient of the cost function over $\Psim$ to a zero matrix, we obtain the update rule for $\Psim$, which is given as
{\small 
\begin{align}
    \frac{\partial \mathcal{L}^{(\ell)}}{\partial \Psim}
    &= Q \Psim^{-1} \!- \!\sum_{i=1}^Q \gamma_i^{-1}\Psim^{-1} \underbrace{\diag(\bar{\tv}_i)^{-1} \Rm_i^{(\ell)} \diag(\bar{\tv}_i^*)^{-1}}_{\triangleq \widetilde{\Rm}^{(\ell)}_i}\Psim^{-1} \!\overset{!}{=}\!\mathbf{0} \notag \\
    &\Rightarrow  \quad \widehat{\Psim}^{(\ell +1)}  = \frac{1}{Q} \sum^Q_{i=1} \left(\widehat{\gamma}_i^{(\ell+1)}\right)^{-1}\widetilde{\Rm}_i^{(\ell)}
\end{align}
}
\section{Illumination optimization}
In this section, we design the illumination pattern by optimizing the BF vectors $\{\xv_n\}$. 
We provide two designs following different criteria. The first design takes into consideration the properties of CS to minimize the total coherence of the sensing matrix so that the basis vectors of the sensing matrix are as mutually orthogonal as possible. The second design maximizes the illumination power of the ROI. We will compare these two designs in the numerical simulation. 
Note that the sensing matrices $\{\Phim_n\}$ of different subcarriers have the same component $\Am$, $\Bm$, $\etav$, and only differ from the BF vector $\xv_n$. Therefore, we will focus on a general subcarrier and drop the subcarrier index $n$ to make the notation simple.   

\subsection{Total Coherence Minimization (TCM)}\label{sec:TCM}
The performance of the CS problem highly depends on the property of the sensing matrix $\Phim$. One of the most fundamental properties is mutual coherence, which represents the worst-case coherence between any two columns of the sensing matrix \cite{ elad2007optimized}. However, mutual coherence can only indicate a worst-case bound and can not reflect the average signal recovery performance \cite{li2013projection}. Additionally, it is also difficult to directly optimize mutual coherence due to its definition of the worst coherence between any two columns. Therefore, we adopt the more commonly considered total mutual coherence as the metric to optimize the illumination pattern. Specifically, we optimize the BF vector $\xv_n$ to minimize the following objective with the power constraint \cite{ge2022training}
\begin{align}
    \min_{\xv} \quad &\|\Phim^\herm \Phim - \alpha^2 \Id_Q\|_{\sfF}^2 \\
    \text{s.t.} \quad &\|\xv\|^2_2 = P,
\end{align}
where $P$ is transmit power, and $\alpha$ is a positive scaling number to satisfy the power constraint.

To solve this problem, we decompose it into three optimization subproblems. First, we directly optimize the sensing matrix $\Phim$ as 
\begin{align}
    \min_{\Phim} \quad &\|\Phim^\herm \Phim - \alpha^2 \Id_Q\|_{\sfF}^2,
\end{align}
which gives the optimized $\Phim$ as \cite{li2013projection}
\begin{align}\label{eq:PHI_star}
     \Phim^\star(\alpha) = \alpha\Um_1 [ \Id_{M_{\rm r}}, \mathbf{0}_{M_{\rm r}\times (Q-M_{\rm r})}]\Um^\herm_2,
\end{align}
where $\Um_1 \in \bC^{M_{\rm r} \times M_{\rm r}}$ and $\Um_2 \in \bC^{Q \times Q}$ are arbitrary unitary matrices. 
Then, let $\betav \triangleq \Bm \xv$, the next subproblem is to solve 
\begin{align}
    \min_{\betav} \quad & f(\betav) = \|\Am \diag(\etav) \diag(\betav) - \Phim \|_{\sfF}^2.
\end{align}
With the expansion of $f(\betav)$ and ignoring the constant term, we have
{\small 
\begin{align}
    f(\betav) &= \trace\left(\Am^\herm \Am \diag(\etav)\diag(\etav)\diag(\betav) \diag(\betav)^\herm\right) \notag\\ 
    &\quad - 2\Re\left\{\trace\left(\Am\diag(\etav) \diag(\betav) \Phim^\herm\right)\right\}
    \\
    &\overset{(a)}{=}\trace\left(\Am^\herm \Am \diag(\eta^2_1|\beta_1|^2,\dots, \eta_Q^2|\beta_Q|^2)\right) 
    \notag\\
    &\quad - 2\Re\left\{\sum_{m=1}^{M_{\rm r}}\sum_{q=1}^{Q} \eta_q \beta_q[\Am]_{m,q}  [\Phim^\herm]_{q,m}\right\} \\
    &=\sum_{q=1}^Q\left([\Am^\herm \Am]_{q,q} \eta_q^2 |\beta_q|^2\right) 
    \notag\\
    &\quad -2\Re\Bigg\{\sum_{q=1}^Q \eta_q\beta_q\underbrace{\left(\sum_{m=1}^{M_{\rm r}} [\Am]_{m,q} [\Phim^\herm]_{q,m}\right)}_{\triangleq g_q   }\Bigg\}\\
    &=\sum_{q=1}^Q \left(\eta_q^2 |\beta_q|^2 - 2 \Re\left\{\eta_q\beta_q  g_q\right\}\right), \label{eq:decouple}
\end{align}
}
where $\beta_q \triangleq [\betav]_q, \eta_q \triangleq [\etav]_q$, $[\Am^\herm \Am]_{q,q} = 1, \; \forall q\in[Q]$, and  the second term in $(a)$ is due to
{\small
\begin{align}
    \left[\Am \diag(\etav)\diag(\betav) \Phim^\herm\right]_{m,m} 
    & = \sum_{q=1}^{Q} [\Am \diag(\etav)\diag(\betav)]_{m,q} [\Phim^\herm]_{q,m} \\ 
    & = \sum_{q=1}^{Q} \eta_q \beta_q [\Am]_{m,q} [\Phim^\herm]_{q,m}. 
\end{align}
}
From \eqref{eq:decouple}, we notice that the variables in $f(\betav)$ are decoupled. Then, we find optimal elements of $\betav$ by setting the partial derivative to zero, which gives 
\begin{align}
    \frac{\partial f(\betav)}{\partial \beta_q} &=  \eta_q^2 \beta_q^* -  \eta_qg_q \overset{!}{=} 0\\
    \Rightarrow \quad \beta_q^\star &= \frac{ g_q^*}{\eta_q}= \frac{[\Phim^\herm \Am]_{q,q}^*}{\eta_q},\label{eq:beta_star}
\end{align}
where $g_q = [\Phim^\herm]_{q,:} [\Am]_{:,q} = [\Phim^\herm \Am ]_{q,q}$. 

Combining steps 1 and 2 by putting \eqref{eq:PHI_star} into \eqref{eq:beta_star}, we have the optimal $\betav^\star(\alpha)$. Finally, the optimal beamforming vector $\xv$ is given by solving 
\begin{align}
    \min_{\xv} \quad &\|\Bm \xv - \betav^\star(\alpha)\|_{2}^2 \\
    \text{s.t.} \quad &\|\xv\|^2_2 = P,
\end{align}
which gives
\begin{align}
    \xv^\star =  (\Bm^\herm \Bm)^{-1} \Bm^\herm \betav^\star(\alpha),
\end{align}
where $\alpha$ is calculated such that $\|\xv^\star\|^2_2 = P$.

\begin{remark}
    Note that if we consider the normalized total coherence, i.e., we use the column-normalized  $\|[\Phim]_{\cdot, q}\|_2 = 1, q\in[Q]$ to calculate the total coherence, then any choice of $\xv$ yields the same result. This is due to the fact that $\xv$ only changes the term of  $\diag(\Bm \xv)$, whose effect will be canceled by column normalization. \hfill $\lozenge$
\end{remark}

\subsection{ Illumination Power Maximization (IPM)}\label{sec:IPM}
In this design, we maximize the minimum illuminated power of all grid points, which results in the following optimization problem in an Epigraph form 
\begin{align}
    \max_{\xv, \chi} \quad & \chi \\
    \text{s.t.} \quad & \eta_i^2|\bv_i^\herm \xv|^2 \geq \chi, \quad \forall i \in [Q], \\
    &\|\xv\|_2^2 \leq P, 
\end{align}
where $\bv_i \triangleq \bv(\rv_i) = [\Bm]_{i, :}^\herm$ is the angular steering vector of the $i$-th cell at the Tx, and $\chi$ is an auxiliary variable. This problem can be equivalently  written as the following semidefinite programming  (SDP)
\begin{align}
    \max_{\Xm\succeq \mathbf{0}, \chi} \quad & \chi \\
    \text{s.t.} \;\quad & \eta_i^2\bv_i^\herm \Xm \bv_i \geq \chi, \quad \forall i \in [Q], \\
    &\trace(\Xm) \leq P, \\
    &\rank(\Xm) = 1.
\end{align}
The problem is non-convex only due to the non-convex rank-1 constraint. We can reform the rank-1 constraint as $\|\Xm\|_* - \|\Xm\|_2 \leq \epsilon$, where $\|\Xm\|_*$ is the nuclear norm of $\Xm$ and $\epsilon$ is a small positive constant very close to zero. In our case with $\Xm\succeq \mathbf{0}$, we have $\|\Xm\|_* = \trace(\Xm)$.  Then, this constraint has a convex-plus-concave form, and the problem can be solved iteratively using the successive convex approximation (SCA) method. 
Specifically, in the $k$-th iteration, the convex term $\|\Xm\|_2$ is  approximated by its lower-bound first-order Taylor expansion   
\begin{align}\label{eq:taylor}
    \|\Xm\|_2 \geq \|\Xm^{(k)}\|_2 + \trace\left(\uv^{(k)} (\uv^{(k)})^\herm (\Xm - \Xm^{(k)})\right),
\end{align}
where $\Xm^{(k)}$ is the feasible point obtained by the previous iteration and $\uv^{(k)}$ is the dominant eigenvector of the $\Xm^{(k)}$ corresponding to the largest eigenvalue. Using this approximation, the resulting problem is a convex SDP problem that can be solved using convex optimization solvers, e.g., CVX \cite{cvx}. 
The initial point $\Xm^{(0)}$ can be obtained by solving the problem without the rank-1 constraint. 
After convergence, the optimal BF vector is obtained as $\xv^\star = \sqrt{P} \uv^{(k)}$.

\begin{remark}
    Note that in the IPM design, the optimized BF vectors for different subcarriers are not necessarily the same. Different cells of the ROI can be taken into account for the optimization at different subcarriers. For instance, in our simulation, we will focus on four disjoint regions of ROI in four subcarriers. By doing so, we can obtain more varied sensing matrices to avoid the bad performance caused by a single low-quality sensing matrix.  
    \hfill $\lozenge$
\end{remark}

\section{Simulation Results}
We consider the topology in Fig.~\ref{fig:IIAC}, where a Tx is located at $(0,20)$ meters and a Rx is located at $(0, -20)$ meters with $M_{\rm t} = M_{\rm r} = 100$ antennas in ULA with half-wavelength antenna spacing, focusing on the square ROI centered at $(0,0)$ under $Q=20\times 20 = 400$ square cells. 
We use $N=4$ subcarriers for imaging with $\Delta f = 1$ MHz frequency spacing under $f_c = 50$ GHz central frequency. We adopt the first-order autoregressive (AR-1) process to model the reflection coefficients at different subcarriers, resulting in a Toeplitz correlation matrix $\Psim = {\sf Toeplitz}([1, \psi, \dots,  \psi^{N-1}])$, where $\psi = 0.99$ is the correlation coefficient. Under such an AR-1 modeling, we can enhance the estimation of $\Psim$ at each iteration by forming the estimation as a Toeplitz matrix with the estimated correlation coefficient being $\widehat{\psi}^{(\ell + 1)} =\frac{m_1}{m_0}$, where $m_0$ and $m_1$ are the average values of the elements along with the main diagonal and main sub-diagonal of $\widehat{\Psim}^{(\ell + 1)}$, respectively \cite{zhang2013extension}.  

We compare three illumination patterns: 1) \textbf{Uniform:} the uniform illumination pattern uses a power-normalized all-ones BF for all subcarriers; 2) \textbf{TCM:} the TCM pattern proposed in Section~\ref{sec:TCM}; 3) \textbf{IPM:} the IPM pattern proposed in Section~\ref{sec:IPM}, in which the optimizations at four subcarriers focus on four different corner parts of the full ROI. 
The true and estimated images are shown in Fig.~\ref{fig:image}, where the imaging results of subcarrier-1 in (b-d) with the corresponding true image in (a) are under high SNR, and the results of subcarrier-4 in (f-h) with the corresponding true image in (e) are under low SNR. 
Additionally, to further evaluate the imaging quality, we consider four commonly used metrics: image mean-squared error (IMMSE), peak signal-to-noise ratio (PSNR), structural similarity (SSIM) index, and Pearson’s correlation
coefficient (PCC).\footnote{Due to the page limit, we omit their details. We use the Matlab functions {\sf immse}, {\sf psnr}, {\sf ssim}, and {\sf corr2} to calculate them. Please refer to the \textit{Image Quality Metrics} in the MATLAB Image Processing Toolbox\texttrademark.} The quantitative results are shown in Table~\ref{tab:result}. 

The pixels of the true image show the short name of our university ``TU Berlin''. The non-zero pixels at subcarrier-1 have gradually decreasing magnitudes from top to bottom and random phases. The true pixels at subcarrier-4 have more random magnitudes due to the AR-1 modeling.
From the numerical results, we first observe that regardless of high or low SNR, the proposed TCM and IPM BF provide much better imaging performance than that of uniform BF, indicating the effectiveness of TCM and IPM. Moreover, under high SNR, TCM performs better than IPM, because TCM can improve the sensing matrices. However, when the SNR is low, TCM has a significant performance loss due to dominant strong noise, while the performance of IPM degrades only slightly, making it more suitable for low SNR cases.

\begin{table}[h!]
\centering
 \begin{tabular}{l | c c c c} 
  & IMMSE & PSNR (dB) & SSIM & PCC \\  
 \hline
 Uniform, high & 0.0849 & 10.71 & 0.6631 & 0.7182 \\ 
 \textbf{TCM, high} & \textbf{0.0046} & \textbf{23.38} & \textbf{0.9841}  & \textbf{0.9862}\\
 IPM, high & 0.0049 & 23.07 & 0.9676  & 0.9852 \\
 \hline
 Uniform, low & 0.1641 & 7.85 & 0.3796 & 0.4703 \\ 
 TCM, low & 0.0155 & 18.09 & 0.8422  & 0.9541\\
 \textbf{IPM, low} & \textbf{0.0055} & \textbf{22.59} & \textbf{0.9185}  & \textbf{0.9839} 
 \end{tabular}
 \caption{Quantitative imaging accuracy comparison of proposed TCM and IPM as well as the baseline Uniform illumination patterns with highlighted best results under high SNR (30 dB) and low SNR (5 dB). }
 \label{tab:result}
 \vspace{-5mm}
\end{table}

\newcommand{\plotsize}{0.22}

\begin{figure*}[t!]
\centering
 \begin{subfigure}[b]{\plotsize\textwidth}
        \includegraphics[width=\columnwidth]{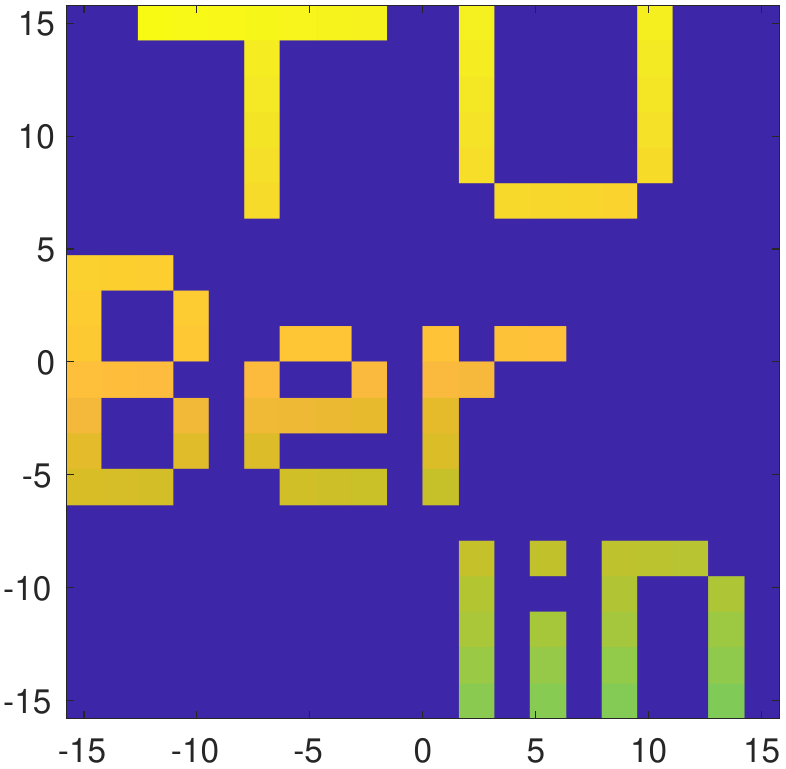}
        \caption{True image subcarrier-1}
        \label{fig:true_1}
        \end{subfigure}
        ~~
        \begin{subfigure}[b]{\plotsize\textwidth}
        \includegraphics[width=\columnwidth]{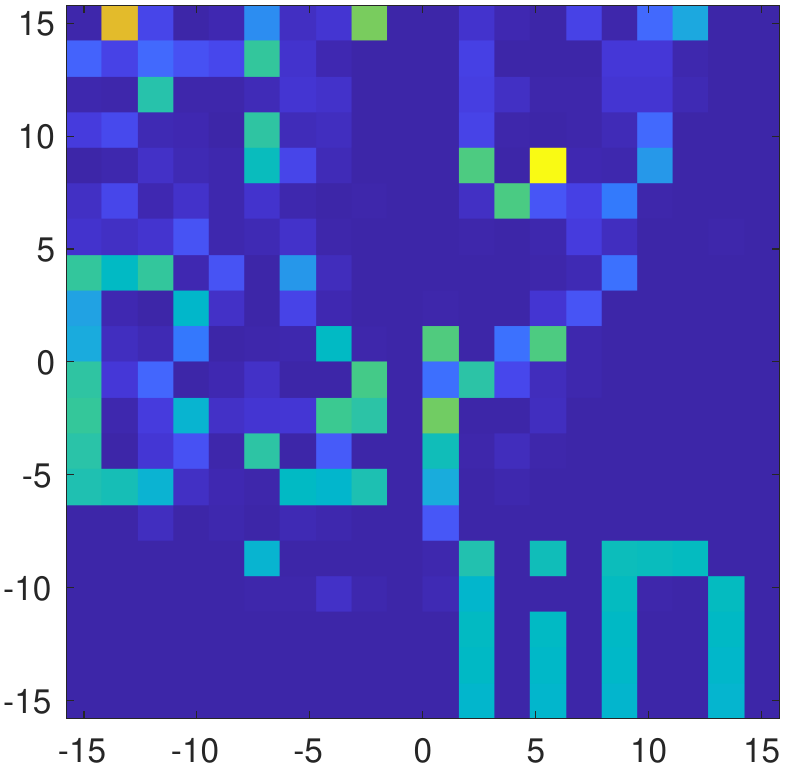}
        \caption{Uniform 30 dB subcarrier-1}
        \label{fig:Uni_1}
        \end{subfigure}
        ~~
        \begin{subfigure}[b]{\plotsize\textwidth}
        \includegraphics[width=\columnwidth]{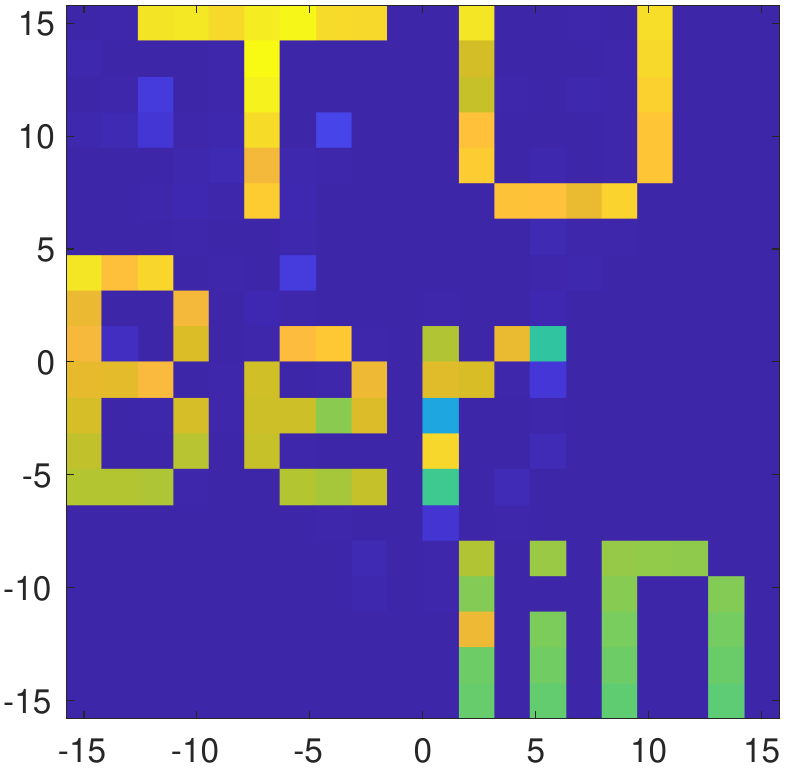}
        \caption{TCM 30 dB subcarrier-1}
        \label{fig:TCM_1}
        \end{subfigure}
        ~~
        \begin{subfigure}[b]{\plotsize\textwidth}
        \includegraphics[width=\columnwidth]{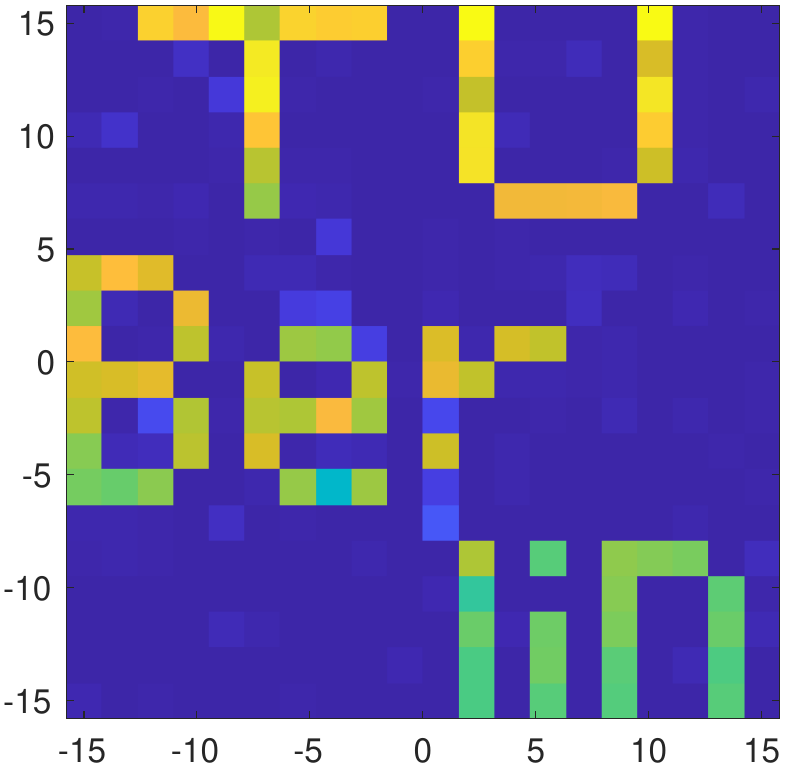}
        \caption{IPM 30 dB subcarrier-1}
        \label{fig:IPM_1}
        \end{subfigure}
        ~~
        \begin{subfigure}[b]{\plotsize\textwidth}
        \includegraphics[width=\columnwidth]{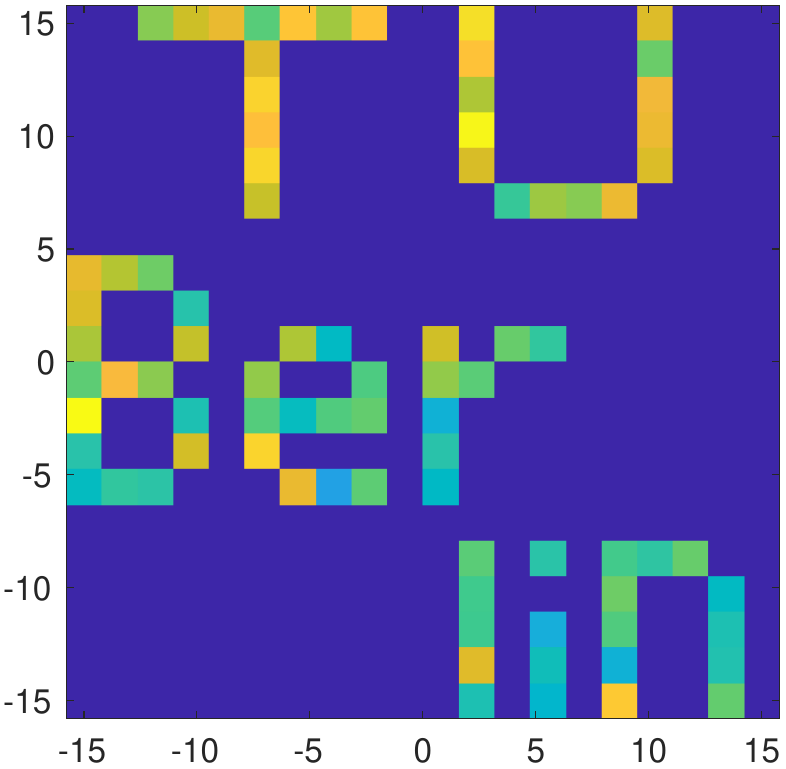}
        \caption{True image subcarrier-4}
        \label{fig:true_4}
        \end{subfigure}
        ~~
        \begin{subfigure}[b]{\plotsize\textwidth}
        \includegraphics[width=\columnwidth]{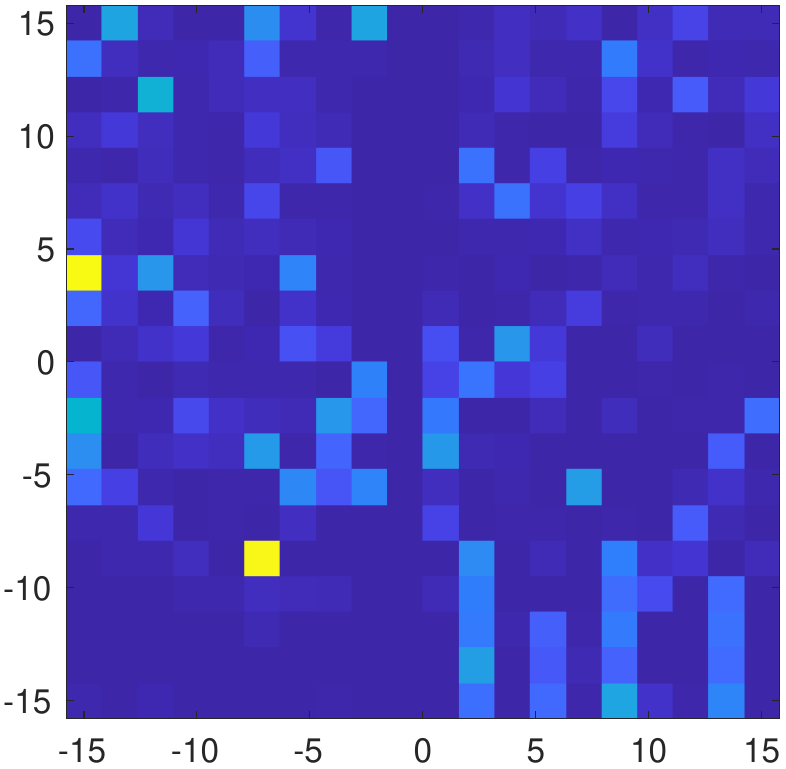}
        \caption{Uniform 5 dB subcarrier-4}
        \label{fig:Uni_4}
        \end{subfigure}
        ~~
        \begin{subfigure}[b]{\plotsize\textwidth}
        \includegraphics[width=\columnwidth]{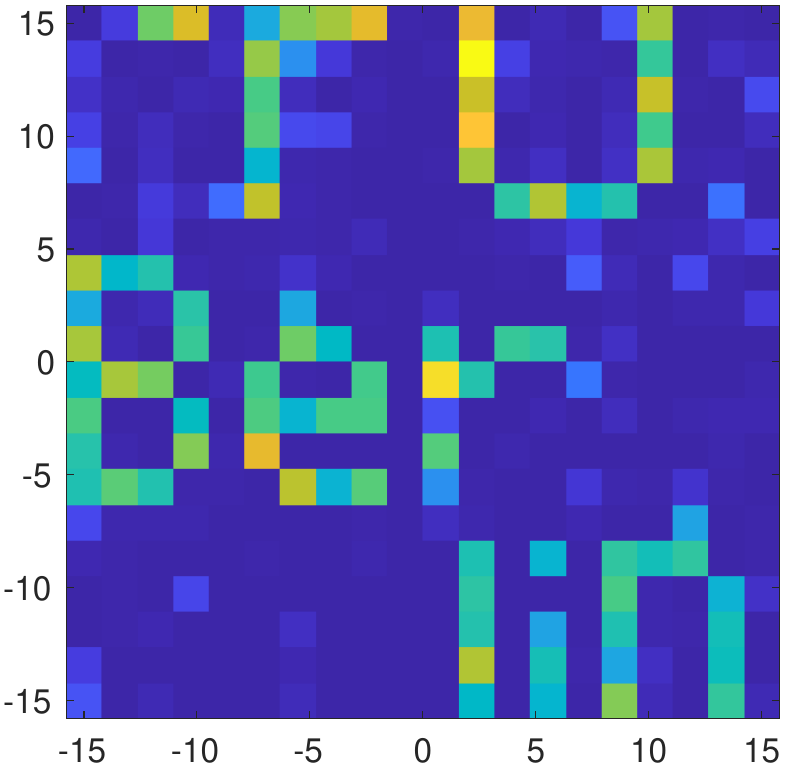}
        \caption{TCM 5 dB subcarrier-4}
        \label{fig:TCM_4}
        \end{subfigure}
        ~~
        \begin{subfigure}[b]{\plotsize\textwidth}
        \includegraphics[width=\columnwidth]{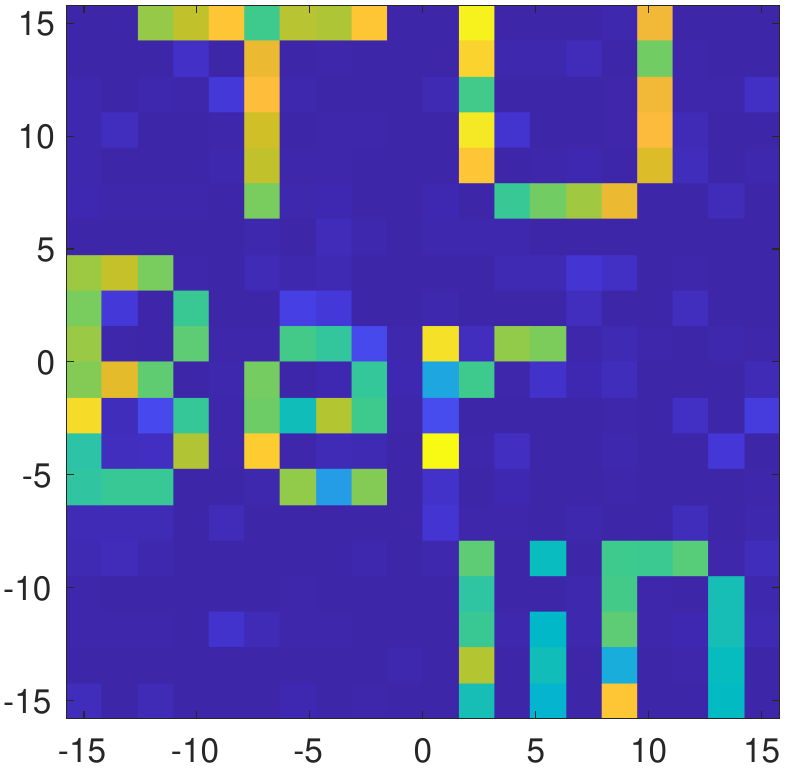}
        \caption{IPM 5 dB subcarrier-4}
        \label{fig:IMP_4}
        \end{subfigure}
        \caption{True and estimated images at subcarrier-1 and subcarrier-4,  (b-d) are under high SNR, (f-h) are under low SNR.}
     \label{fig:image}
     \vspace{-5mm}
\end{figure*}

\newpage

\section{Conclusion}
In this paper, we addressed the imaging problem using wideband and near-field communication signaling. In order to exploit the frequency correlation of reflection coefficients, we formulated the imaging as a special block-sparse MMV problem with correlation of observations and various sensing matrices. We proposed an SBL-based approach that alternatively estimates both the image coefficients and their correlation in frequency. We also optimized two illumination patterns separately to improve the CS performance and SNR level.
The numerical results showed the excellent performance of the proposed SBL-based imaging solution and the effectiveness of the optimized illumination patterns.

{\small
	\bibliographystyle{IEEEtran}
	\bibliography{references}
}

\end{document}

%% file: symbols.tex
\usepackage{mathbbol}

\def\mindex#1{\index{#1}}

\def\sq{\hbox{\rlap{$\sqcap$}$\sqcup$}}
\def\qed{\ifmmode\sq\else{\unskip\nobreak\hfil
\penalty50\hskip1em\null\nobreak\hfil\sq
\parfillskip=0pt\finalhyphendemerits=0\endgraf}\fi\medskip}

\long\def\defbox#1{\framebox[.9\hsize][c]{\parbox{.85\hsize}{%
\parindent=0pt
\baselineskip=12pt plus .1pt      
\parskip=6pt plus 1.5pt minus 1pt 
 #1}}}

\long\def\beginbox#1\endbox{\subsection*{}%
\hbox{\hspace{.05\hsize}\defbox{\medskip#1\bigskip}}%
\subsection*{}}

\def\endbox{}

\def\diag{{\text{diag}}}

\def\rank{{\rm rank\,}}

\newsavebox{\junk}
\savebox{\junk}[1.6mm]{\hbox{$|\!|\!|$}}

\def\det{{\mathop{\rm det}}}

\def\Re{\field{R}}

\def\bC{{\mathbb C}}

\def\bE{{\mathbb E}}

\def\bR{{\mathbb R}}

\def\sfF{{\sf F}}

\def\bfmath#1{{\mathchoice{\mbox{\boldmath$#1$}}%
{\mbox{\boldmath$#1$}}%
{\mbox{\boldmath$\scriptstyle#1$}}%
{\mbox{\boldmath$\scriptscriptstyle#1$}}}}

\def\bfmY{\bfmath{Y}}

\def\bfmhhaY{\bfmath{\hhaY}} 
\def\bfmhhaY{\hbox to 0pt{$\widehat{\bfmY}$\hss}\widehat{\phantom{\raise 1.25pt\hbox{$\bfmY$}}}}

\def\til={{\widetilde =}}

 \def\FRAC#1#2#3{\genfrac{}{}{}{#1}{#2}{#3}}

\def\ddtp{{\mathchoice{\FRAC{1}{d^{\hbox to 2pt{\rm\tiny +\hss}}}{dt}}%
{\FRAC{1}{d^{\hbox to 2pt{\rm\tiny +\hss}}}{dt}}%
{\FRAC{3}{d^{\hbox to 2pt{\rm\tiny +\hss}}}{dt}}%
{\FRAC{3}{d^{\hbox to 2pt{\rm\tiny +\hss}}}{dt}}}}

\def\average#1,#2,{{1\over #2} \sum_{#1}^{#2}}

\def\eye(#1){{\bf(#1)}\quad}

\newtheorem{remark}{{\bf Remark}}

\def\eq#1/{(\ref{e:#1})}

\newcommand{\beqn}[1]{\notes{#1}%
\begin{eqnarray} \elabel{#1}}

\newcommand{\eeqn}{\end{eqnarray} }

\newcommand{\beq}[1]{\notes{#1}%
\begin{equation}\elabel{#1}}

\newcommand{\eeq}{\end{equation}}

\def\bdes{\begin{description}}
\def\edes{\end{description}}

\newcounter{rmnum}

\newcounter{anum}

{\end{list}}

\def\ass(#1:#2){(#1\ref{#1:#2})}

\def\ritem#1{
\item[{\sf \ass(\current_model:#1)}]
}

\newenvironment{recall-ass}[1]{%
\begin{description}
\def\current_model{#1}}{
\end{description}
}



%% file: macros.tex
\long\def\comment#1{}

\newfont{\bb}{msbm10 scaled 1100}

\newcommand{\av}{{\bf a}}
\newcommand{\bv}{{\bf b}}

\newcommand{\nv}{{\bf n}}

\newcommand{\rv}{{\bf r}}

\newcommand{\tv}{{\bf t}}
\newcommand{\uv}{{\bf u}}

\newcommand{\xv}{{\bf x}}
\newcommand{\yv}{{\bf y}}

\newcommand{\Am}{{\bf A}}
\newcommand{\Bm}{{\bf B}}

\newcommand{\Hm}{{\bf H}}
\newcommand{\Id}{{\bf I}}

\newcommand{\Nm}{{\bf N}}

\newcommand{\Pm}{{\bf P}}

\newcommand{\Rm}{{\bf R}}

\newcommand{\Tm}{{\bf T}}
\newcommand{\Um}{{\bf U}}
\newcommand{\Wm}{{\bf W}}

\newcommand{\Xm}{{\bf X}}
\newcommand{\Ym}{{\bf Y}}
\newcommand{\Zm}{{\bf Z}}

\newcommand{\betav}{\hbox{\boldmath$\beta$}}
\newcommand{\gammav}{\hbox{\boldmath$\gamma$}}

\newcommand{\etav}{\hbox{\boldmath$\eta$}}

\newcommand{\muv}{\hbox{\boldmath$\mu$}}

\newcommand{\rhov}{\hbox{\boldmath$\rho$}}

\newcommand{\Gammam}{\hbox{\boldmath$\Gamma$}}

\newcommand{\Sigmam}{\hbox{\boldmath$\Sigma$}}
\newcommand{\Phim}{\hbox{\boldmath$\Phi$}}

\newcommand{\Psim}{\hbox{\boldmath$\Psi$}}

\renewcommand{\det}{{\hbox{det}}}
\newcommand{\trace}{{\hbox{tr}}}

\renewcommand{\Re}{{\rm Re}}

\newcommand{\herm}{{\sf H}}

\newcommand{\transp}{{\sf T}}
\renewcommand{\vec}{{\rm vec}}

